\documentclass[prl,twocolumn,preprintnumbers,amsmath,amssymb,floatfix,reprint]{revtex4}
\usepackage{graphicx}
\usepackage{epsfig}
\usepackage{color}
\begin{document}

\title{Nanofriction in Cold Ion Traps}

\author{A. Benassi$^{1,2}$, A. Vanossi$^{2,1}$ and E. Tosatti$^{2,3,1}$}
\affiliation{
$^1$ CNR-IOM Democritos National Simulation Center,Via Bonomea 265, I-34136 Trieste, Italy \\
$^2$ International School for Advanced Studies (SISSA), Via Bonomea 265, I-34136 Trieste, Italy \\
$^3$ International Centre for Theoretical Physics (ICTP), P.O.Box 586, I-34014 Trieste, Italy
}


\maketitle

\section{abstract}
Sliding friction between crystal lattices and the physics of cold ion traps are so far non-overlapping fields.
Two sliding lattices may either stick and show static friction or slip with
dynamic friction; cold ions are known to form static chains, helices, or clusters, depending on
trapping conditions. Here we show, based on simulations, that much could be learnt about friction by
sliding, via e.g. an electric field, the trapped ion chains over a periodic corrugated potential. Unlike infinite
chains where, according to theory, the classic Aubry transition to free sliding may take place, trapped chains are always pinned.
Nonetheless we find that a properly defined static friction still vanishes Aubry-like at a
symmetric-asymmetric structural transition, ubiquitous for decreasing corrugation in both straight and
zig-zag trapped chains. Dynamic friction can also be addressed by ringdown oscillations of the ion trap.
Long theorized static and dynamic one dimensional friction phenomena could thus become exquisitely accessible
in future cold ion tribology.

\section{Introduction}
The field of sliding friction has been recently revived thanks to experimental and theoretical advances
especially connected with nanosystems, with brand new opportunities to grasp the underlying phenomena
of this important and technologically relevant area. Unlike macroscopic friction,
where contact irregularities dominate, the nanoscale offers perfectly defined crystal lattice facets that can
mutually slide. One fundamental piece of physics arising in this limit is the possibility of frictionless sliding when
two perfect lattices are mutually incommensurate -- their periodicities unrelated through any rational number--
an idealized situation sometimes referred to as superlubric~\cite{hirano}.
Graphite flakes sliding on graphite were for example shown to be pinned by static friction -- the finite
threshold force $F_s$ necessary to provoke motion -- when aligned, but to turn essentially free sliding
once a rotation makes them incommensurate ~\cite{dienwiebel}.
An additional prediction, rigorously proven for one dimensional (1D) infinite harmonic chains sliding
in a periodic potential (the ``corrugation'' potential, in frictional language), is that the
incommensurate vanishing of static friction will only occur so long as the sliding chain is sufficiently
hard compared to the corrugation strength. When instead that strength exceeds a critical value,
the chain becomes locked, or ``pinned'', to the corrugation, thus developing static friction despite
incommensurability. Unlike most conventional phase transitions, this celebrated Aubry transition~\cite{aubry,braun}
involves no structural order parameter and no breaking of symmetry -- just a change of phase
space between the two states, the unpinned and the pinned one. The concept of pinned and unpinned
states is by now qualitatively established in the sliding of real incommensurate 3D crystal surfaces~\cite{persson}.
Yet, the onset of static friction in the prototypical 1D chain, where the Aubry transition is mathematically established, has
never been experimentally validated. Even less is known about dynamic friction, about which we have no
other insight than generic linear response~\cite{cieplak}, and some data on 3D rare gas overlayers~\cite{coffey}
both suggesting a quadratic increase with corrugation.

\begin{figure}
\epsfig{file=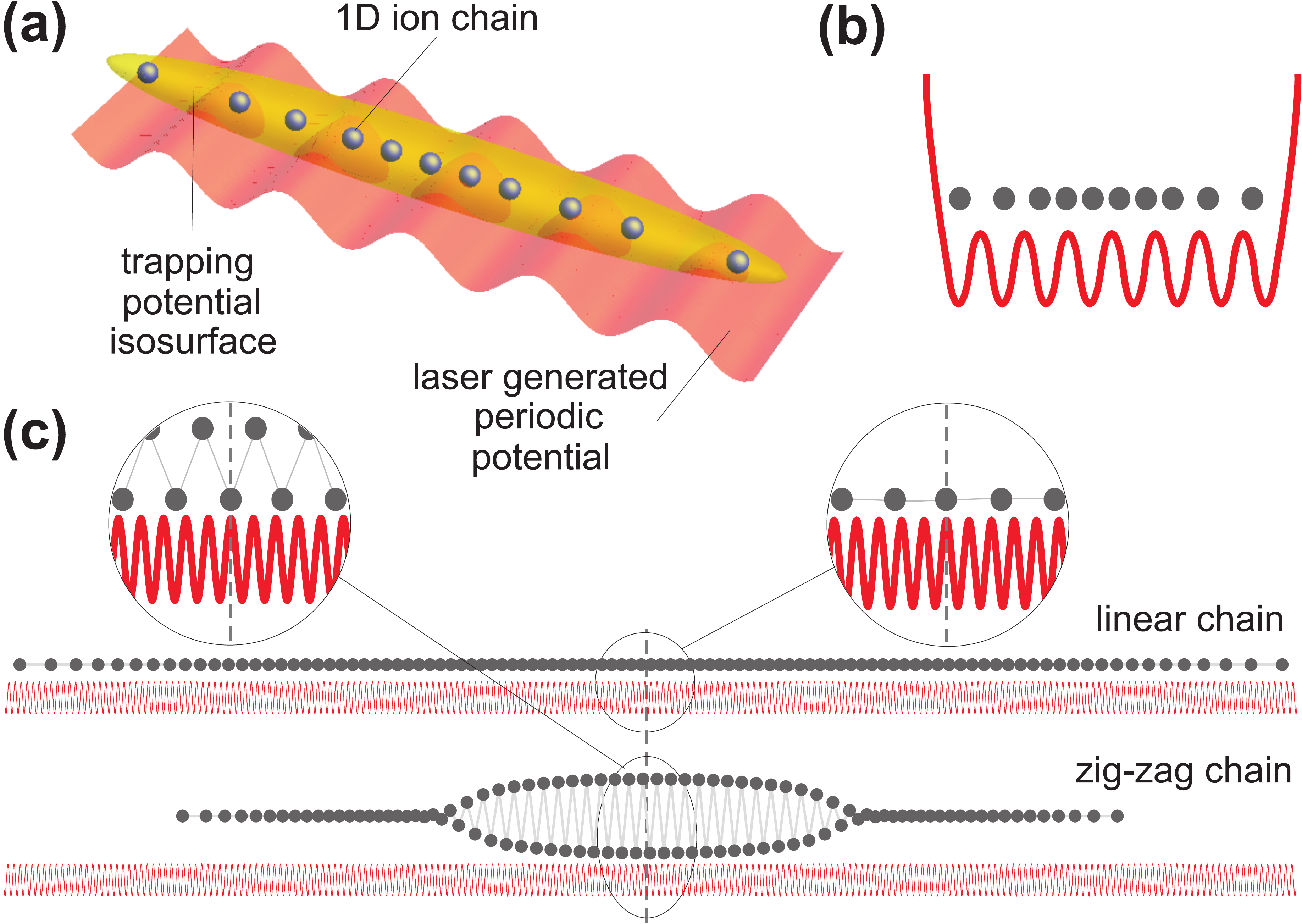,width=8.5cm,angle=0}
\caption{\label{figura1} {(a) and (b) sketch of proposed ion trap configuration in a periodic potential;
(c) sketch of the straight and zig-zag chain equilibrium geometries determined by different aspect ratios of the trapping quadrupolar potential. The
lateral zig-zag amplitude has been artificially magnified.}}
\end{figure}
Cold atomic ions can form linear chains. Despite their mutual Coulomb repulsion, ions can be corralled
inside effective potential traps generated by quadrupolar radio-frequency electrodes~\cite{ghosh}. The
low temperature ion equilibrium geometry is determined by the aspect ratio asymmetry of the confining potential, according
to the effective hamiltonian~\cite{morigi,morigi2}
\begin{equation}
\label{hamiltonian}
H_0=  \sum_i \bigg\{\frac{1}{2}m\dot {\bf r}_i ^2 + \sum\limits_{j \ne i}^N {\frac{(e^2}{{\left| {\mathbf{r}_i  - \mathbf{r}_j } \right|}}} + \frac{1}{2}m\left[ {\omega _ \bot ^2 (x_i^2  + y_i^2 ) + \omega _\parallel ^2 z_i^2 } \right]  \bigg\}
\end{equation}
where $\mathbf{r}_i=(x_i,y_i,z_i)$ is the position of the $i$-th ion, $m$ its mass, $e$ its charge, and the three terms
represent kinetic energy, Coulomb repulsion, and the effective parabolic trapping potential respectively,
with transverse and longitudinal trapping frequencies $\omega_{\perp}$ and $\omega_{\parallel}$.
As is known experimentally~\cite{raizen} and theoretically~\cite{schiffer,fishman}, by decreasing the
asymmetry aspect ratio $R$ = $\omega _\parallel  / \omega _ \bot$ the trap potential deforms from
spherical to cigar-shaped, and the equilibrium ion geometry changes from 3D clusters to helices to chains,
initially zig-zag and finally linear and straight along $z$, as in Fig.~\ref{figura1}(c). As many as a hundred
ions may be stabilized in a linear configuration, with a few $\mu$m ion-ion spacing $a_0$,
typical repulsion energy $e^2/a_0\sim 0.5$ K, and temperatures below $\sim 1 \mu$K.
So far, cold ion chains raised interest in view of promising applications for spectroscopy~\cite{kienle,Hermanspahn,Koerber},
frequency standards~\cite{Bize,Poitzsch}, study and control of chemical reactions~\cite{Molhave},
and quantum information~\cite{Cirac,Steane,Schmidt,Leibfried}.
We are concerned here with the possibility that they could be of use in the field of friction.
An interesting observation was made by Garcia-Mata et al.~\cite{mata}, that ion chains in an additional
incommensurate periodic potential (produced, e.g., by suitable laser beams) would, if only the spatial inhomogeneity
of the trapping potential could be neglected, resemble precisely the 1D system where an Aubry transition
is expected as a function of corrugation strength. Unfortunately, the trapping potential itself
introduces an additional and brutal break of translational invariance, at first sight spoiling this neat idea.
The ion-ion spacing is only uniform near the chain center, increasing strongly towards the two chain ends,
where it diverges owing to the trap confinement (see e.g., Fig.~\ref{figura1}).
Incommensurability of the chain with the periodic potential is therefore lost, and seemingly with that all
possibilities to study phenomena such as the vanishing of static friction at the Aubry transition, a transition
known to depend critically on a precisely defined value of incommensurability~\cite{mukamel}.\\

\begin{figure}
\epsfig{file=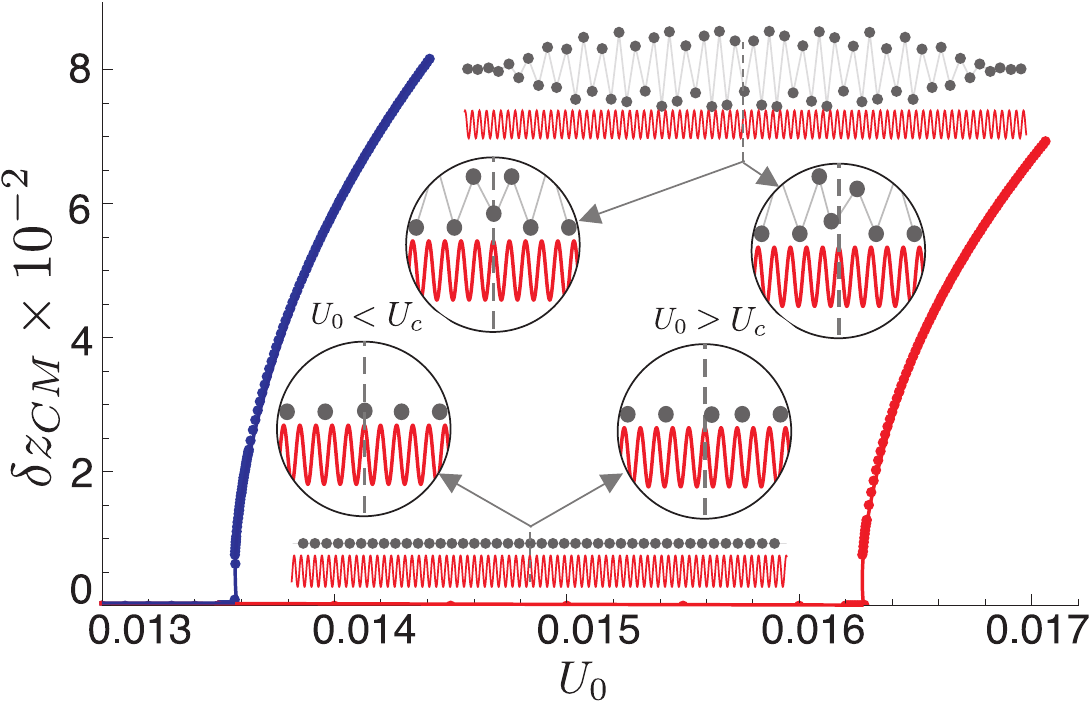,width=8.5cm,angle=0}
\caption{\label{figura2} {Zero temperature ion chain equilibrium center-of-mass coordinate $\delta z_{CM}$ versus periodic corrugation
amplitude $U_0$ for the linear chain (red) and zig-zag chain (blue). Position, time and energies in dimensionless units (see methods).
The insets zoom on the linear chain and the zig-zag central portions, highlighting the symmetry breaking
taking place at $U_0=U_c$ and above.}}
\end{figure}

However, this conclusion is overly pessimistic. In contexts unrelated to friction, a finite
chain in a periodic potential~\cite{sharma,furuya,braiman} is known to turn the would-be Aubry transition to a
structural phase transition. The transition survives, as we will show, despite the addition of the confining
trap potential. The distorted state, prevailing at strong corrugation, exhibits under
suitable conditions a change of symmetry and an accompanying displacive order parameter,
measuring the breaking of inversion symmetry about the chain center.
Simulations carried out by pulling the trapped chain with a longitudinal force (such as in principle
provided by a static external electric field) reveal that the distortion transition
is associated with a continuous onset of static friction from zero to finite, thus mimicking at finite size, and
despite the trapping potential, the static friction onset theoretically expected at the ideal Aubry transition.

In the rest of this work explicitly simulated {\it gedanken experiments} will demonstrate that
trapped cold ions may in fact represent an idesal system for future nanotribological studies addressing
both static and dynamic friction. In static friction, the celebrated Aubry transition will become 
experimentally accessible for the first time. In dynamics, the friction rise with increasing corrugation
should become measurable giving new impulse to future theory work. 

\section{Results}

\subsection{Model and symmetry breaking transition.}
We simulate by means of standard (weakly damped) classical molecular dynamics $N=101$ ions trapped
by the parabolic potential of Eq.(\ref{hamiltonian}), and seek first the static ($T=0$) equilibrium
geometry. Below a critical aspect ratio $R_0=1.18\cdot 10^{-3}$ the equilibrium geometry is a linear chain,
while at $R=R_0$ there is a transition to a planar partly zig-zag structure (see Fig.~\ref{figura1} and Supplementary Movie 1),
further evolving to a helical structure, and eventually to a 3D cluster when $R \rightarrow 1$~\cite{schiffer}.
Choosing $R=0.5\cdot 10^{-3}$ for the frictional simulation of 1D trapped linear chain, we introduce an additional
periodic ``corrugation'' potential $U =  U_0 \cos \left( {\frac{{2\pi }}{\lambda }z} \right)$. To address
the best studied case of Aubry transition, we choose the corrugation wavelength $\lambda$ to be
``golden ratio'' incommensurate relative to the chain center ion-ion spacing $a_0$, $\lambda=2 a_0/(1+\sqrt{5})$
~\cite{aubry,braun}. Preserving the chain's inversion symmetry, the periodic potential phase is chosen so
that the central ion sits at a potential maximum. At vanishing external force, the chain equilibrium configuration
can then be followed while adiabatically increasing the corrugation amplitude $U_0$. As shown in Fig.~\ref{figura2},
the chain ground state remains fully left-right symmetric right up to a critical corrugation amplitude $U_c=0.01627$,
(in dimensionless units, see Methods), where it sharply turns asymmetric, the center
of mass (CM) moving right or left by an amount $\delta z_{CM}(U_0)$ (see Supplementary Movie 2). This transition, similar but not identical
to that known for free finite chains~\cite{sharma,braiman} is the remnant of the Aubry transition
of the ideal infinite chain. In both the finite and the infinite chain, the probability to find an ion (here the central ion)
sitting exactly at a periodic potential maximum vanishes above the critical corrugation strength, where the
ion moves out to a minimum distance $\psi$ away from the potential maximum~\cite{coppersmith,braun}. Unlike
the Aubry transition that has no structural order parameter, the ($T=0$) finite chain transition is
structural, characterized by a displacive order parameter $\delta z_{CM}$, Ising-like since
the central ion can identically fall left or right of the centre. This order parameter grows
continuously as $|U_0-U_c|^\beta$ at the transition, (see Fig.~\ref{figura2}). We find $\beta = 0.496 \pm 0.005$
for the 
trapped 
linear
ion 
chain -- practically coincident with the exponent $\kappa = 0.5$ known for finite chains
with short range forces and without trapping potential~\cite{braiman}. Moreover, upon reducing
the confining shape asymmetry by increasing the aspect ratio to $R=1.2\cdot 10^{-3}$, the chain equilibrium
geometry evolves from linear to a partly planar zig-zag (Fig.~\ref{figura2}), still symmetric about $z=0$
at $U_0=0$. For increasing corrugation a symmetry-breaking transition similar to that of the linear chain
takes place here too at $U_c=0.01357$ -- a smaller value since in the zig-zag the ions are further apart --
and for such case $\beta=0.504 \pm 0.005$ (see Supplementary Movie 3). For both the linear and the 
zig-zag chain the critical amplitude $U_c$ also depends upon the incommensurability ratio $\lambda/a_0$ (not shown).
While these results are valid at $T$=0, it should be stressed that as in all small size systems,
temperature will cancel all symmetry breaking and order parameters, owing to thermal jumps taking
place above the energy barrier $\Delta$ which separates the two opposite Ising like order parameter
valleys.
However, the mean time lapse between jumps diverges exponentially when $T << T_b \sim  \Delta / k_B$,
the trap blocking temperature. Except too close to $U_c$, ion traps are routinely cooled down to temperatures of order $10^{-7}$
(in our units $e^2/d \sim 6.3$ K) way below the blocking temperature (for example $T_b \sim 180 \mu$K even for $U_0 = 0.014$,
barely above the critical $U_c=0.01357$). 
Even at finite temperatures therefore, 
the ion chain will 
still break inversion symmetry and
exhibit a distortive order parameter for macroscopically long times similar to the $T=0$ results
detailed above.

\subsection{Determination of the static friction force.}
\begin{figure}
\epsfig{file=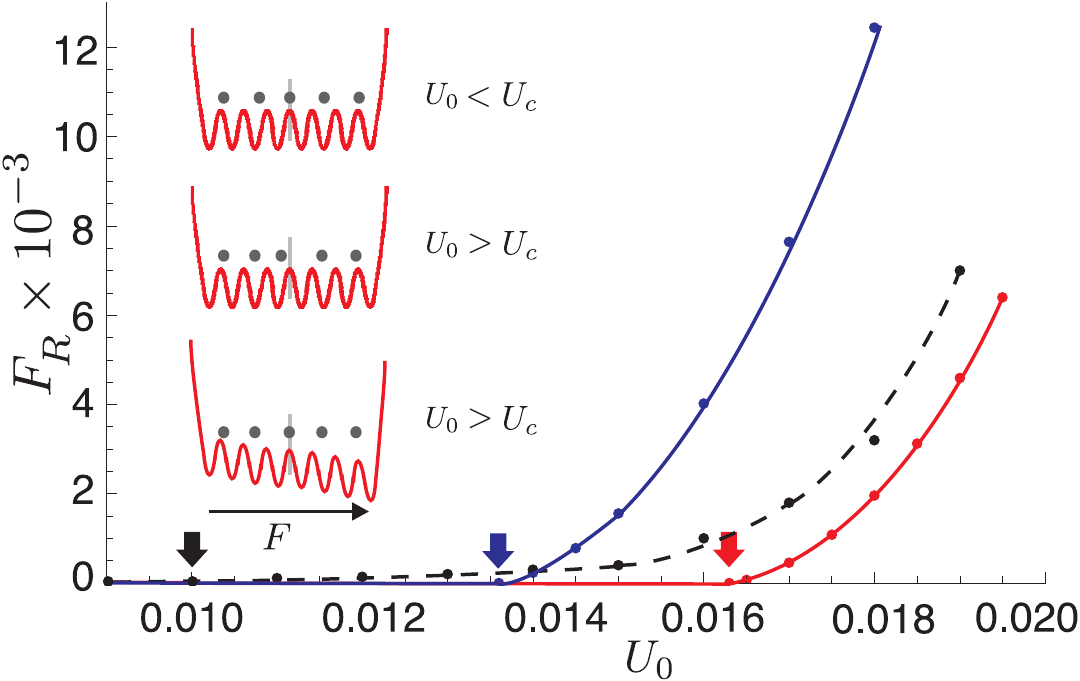,width=8.5cm,angle=0}
\caption{\label{figura3} {Effective static friction measured by the restoring force $F_R$
necessary to symmetrize the system about the center, versus corrugation amplitude $U_0$ for the trapped linear ion chain (red)
and zig-zag chain (blue). The insets illustrate the procedure used. Black dashed line:
calculated theoretical static friction at the Aubry transition of the ideal infinite golden
ratio incommensurate chain. Arrows mark the vanishing point of effective static friction
in each case.} }
\end{figure}

Thus far the description of the ion trap in a periodic corrugation, while still free of external forces. To study
static friction we apply a uniform force $F$, 
such as that of
an electric field parallel to the chain axis. The usual procedure
used to determine static friction -- the smallest force capable of causing
a pinned-to-sliding transition, following which the chain is depinned with unlimited sliding -- does not work here,
since the field-induced center of mass displacement $\delta z_{CM}(F)$ is finite by necessity, owing to the
confining trap potential. We circumvent that problem by monitoring another quantity, namely the
restoring force $F_R$ that must be applied to the chain in order to shift its CM coordinate back exactly to zero --
keeping the chain's central ion exactly on top of the corrugation maximum at $z=0$. In the infinite free chain
this alternative definition of static friction coincides with the standard pinned-to-sliding Peierls-Nabarro
threshold force $F_{PN}$~\cite{peyrard,braun}, identifiable with the static friction $F_s$ of the
system. In the pinned Aubry state the critical force $F_{PN}$ 
is exactly such as to 
drive to zero the minimum equilibrium
deviation $\psi$ of one ``kink" away from the nearest corrugation maximum, eventually triggering the free motion
which initiates the pinned-to-sliding transition of the whole system~\cite{braun_paper,braun}. Analogously,
for trapped ions a growing external force 
$F$
gradually causes the central atom deviation from the potential maximum
to decrease, and eventually to vanish at $F=F_R$. We conclude that $F_R$ measures the static friction even in a
trap, where proper free sliding is impeded. As shown in Fig.\ref{figura3} (inset), for $U_0<U_c$ the central ion
of the trap spontaneously remains precisely at zero, whence $F_R=0$. However at $U_0>U_c$ where the
force-free 
displacive order parameter $\delta z_{CM}$ turns spontaneously nonzero, a finite restoring
force $F_R$ must be applied in order to cancel it.
The growth of the symmetry restoring force $F_R$ with $U_0$ ( Fig.~\ref{figura3}) is, despite the
larger threshold, generally close to the Aubry static friction expected theoretically for the infinite ideal chain~\cite{aubry,hirano} ,
when calculated for same chain parameters $a_0$, $U_0$ and golden ratio $\lambda$, as those at the trapped chain center. 
We have thus established a connection between the trapped chain restoring force
$F_R$ and the Peierls-Nabarro static friction of the infinite chain, of tribological significance~\cite{hirano}.
Numerical differences between the two are of course hardly surprising. The finite size $Na_0$ cuts off the Aubry
critical region where the correlation length exceeds this size. For the same reason the chain depinning exponent
$\chi$, $F_R \sim |U_0 - U_c|^{\chi}$, which we find to be about $1.85 \pm 0.2$ differs from the higher exponent $\sim 3$
expected for the infinite free chain for short range forces~\cite{peyrard,coppersmith,braun}. The effective static
friction of linear and zig-zag chains also differ slightly (see Fig.~\ref{figura3}), the softer zig-zag
chain easier to pin, owing to its larger compliance than the linear chain.
Overall, both cases provide a measurable realization of the sharp onset of static friction at the
Aubry transition, valid beyond the limitations of a trapped chain.

\subsection{Dynamic friction.}
Besides static friction and its Aubry-like vanishing transition, ion traps could 
in addition provide nontrivial insight on dynamic friction. The onset of dynamical friction in depinned
infinite chains is a very subtle event as discussed in literature~\cite{consoli}; 
the pinned-to-sliding onset has been simulated and elegantly described by Braun and coworkers~\cite{braun_paper,braun}. 
In tribology we are primarily interested in the growth, for incommensurate sliders at finite sliding speed, 
of dynamical friction with increasing corrugation.

\begin{figure}
\epsfig{file=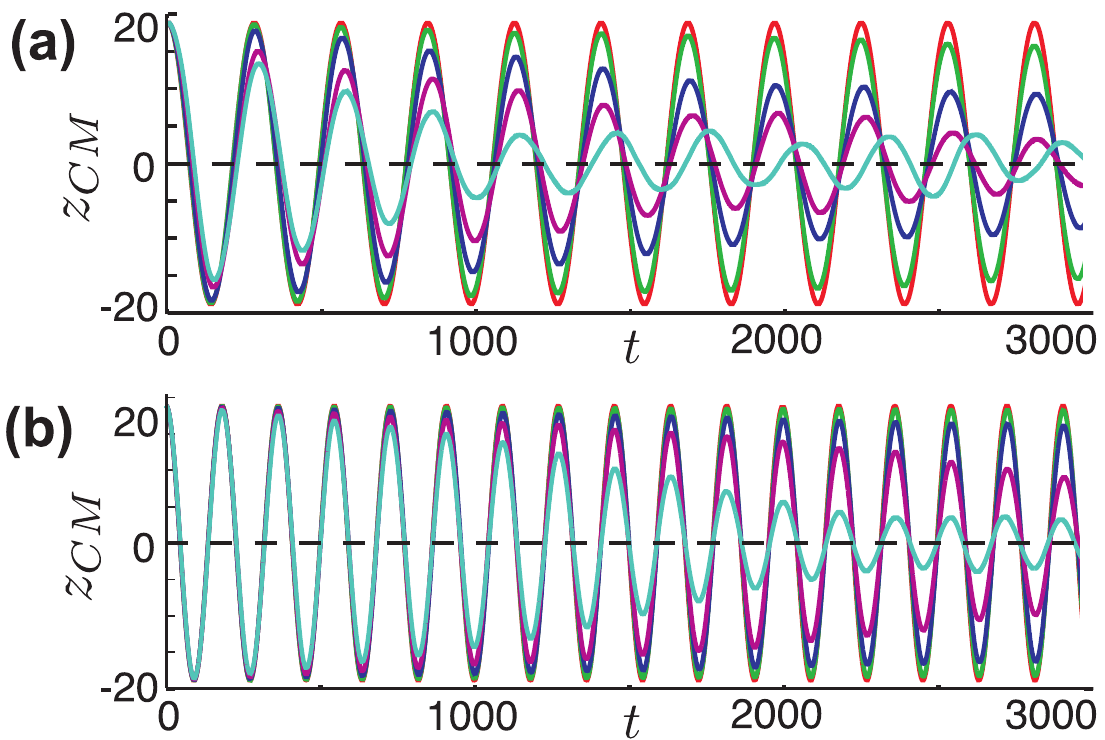,width=8.5cm,angle=0}
\caption{Oscillatory kinetic ringdowns for the linear (upper panel) and the zig-zag (lower panel) chains. The two panels show the $z_{CM}$ coordinate as a function of time for different values of $U_0<U_c$: red $U_0=0$, green $U_0=0.0001$, blue $U_0=0.0002$, magenta $U_0=0.0003$ and cyan $U_0=0.0004$.}
\label{figura4}
\end{figure}

\begin{figure}
\epsfig{file=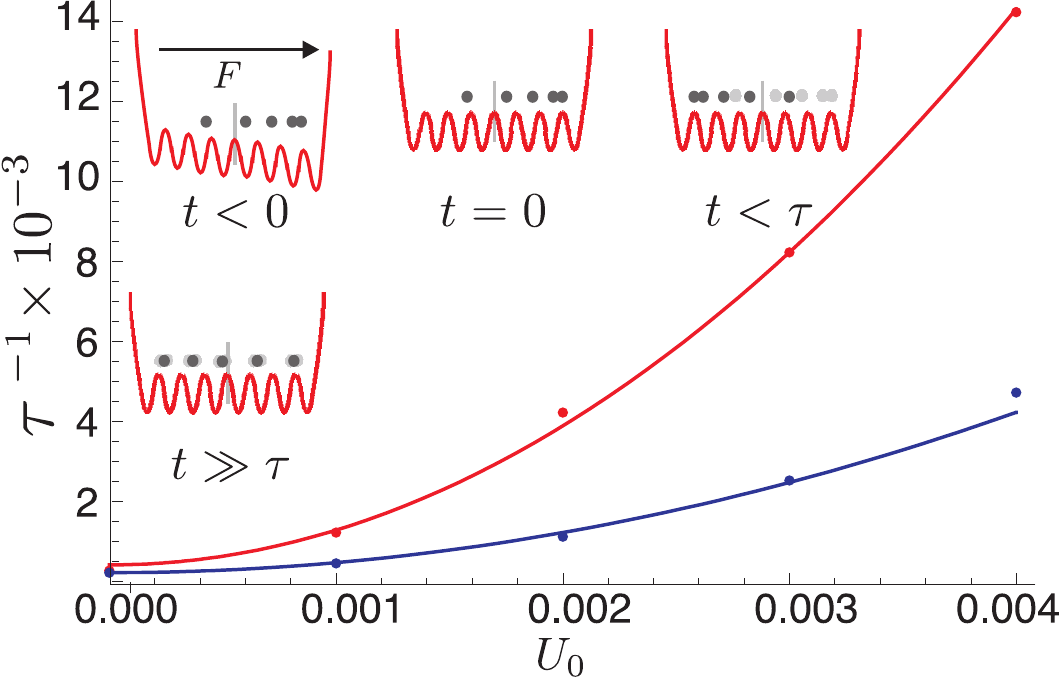,width=8.5cm,angle=0}
\caption{\label{figura5} {(a) Inverse decay time $\tau^{-1}$ of CM ringdown oscillations, measuring the
dynamic friction of the ion chain, linear (red) and zig-zag (blue). The inset illustrates the procedure
used to obtain it. Continuos lines represent fits of the form $a+bU_0^2$ ($a=0.014$ and $b=867.82$ for
the linear chain and $a=0.0$ and $b=250.31$ for the zig-zag).}}
\end{figure}

To simulate the chain dynamical behaviour we initially displace the ions away from their
equilibrium configuration with a static field, then remove the field and follow the spontaneous ``ringdown'' damping
of the bodily oscillation of the whole chain in the trap potential. For an initial large CM displacement of
order $10$ $a_0$ the chain oscillation damping is readily observable, and can for weak corrugation
$U_0 \ll U_c$ be measured even in the absence of an external heat bath. The
initial potential energy (here about 0.1) is converted to Joule heating of the chain, at a ringdown rate which is a
direct measure of dynamic friction. Actually, a trapped chain has an intrinsic dynamic friction even in the absence
of corrugation, because of the trap-potential-induced transfer of kinetic energy from the CM motion to
internal chain phonon modes. However, this effect is small; the conversion and the corresponding dynamic
friction is strongly enhanced by the corrugation potential, as shown in Fig.~\ref{figura4} and in Supplementary Movies 4 and 5 
(for the linear and zig-zag configurations of the chain, respectively). Simulation of the chain oscillations
yields a ringdown amplitude decay for both linear and zig-zag chains,
of the form $A_0 exp-[t/\tau]$. We find that the lifetime $\tau (U_0)$ strongly decreases for increasing $U_0$.
The corresponding increase of dynamic friction $F_D$, directly proportional to $\tau^{-1}$, is to a good
approximation quadratic, $F_D \sim a+bU_0^2$ for weak corrugation (Fig.~\ref{figura5}). A quadratic increase is just what
the linear response theory predicts based on Fermi's golden rule~\cite{cieplak,liebsch}. It also agrees
with experimental Quartz Crystal Microbalance monolayer friction data of Coffey and Krim~\cite{coffey}.
There is in the literature no dynamic friction theory for the infinite linear chain to be used for comparison
with the trapped chain results of Fig.~\ref{figura5}; the present results can therefore be considered a first
exploratory result in this direction. The exceedingly small values of $a$ speak for the strong harmonicity of the chain,
whose CM motion can only with great difficulty excite the higher frequency intra-chain phonons in the absence
of corrugation. The quadratic increase with $U_0$ reflects the effective opening of anharmonic excitation
channels through Umklapp scattering. Larger oscillation amplitudes
could not be readily simulated because the Joule heat liberated would in that case be excessive and
would destroy the chain in the lack of some external dissipation mechanism. A fully nonlinear behavior
of dynamic friction would be expected to emerge in that case, with outcomes strongly
dependent upon the heat disposal mechanism.

\section{Discussion and Conclusions}

The remarkable static and dynamic friction properties emerging for trapped ion chains
seem well amenable to experimental study. The exceedingly small values of the distortion $\delta z_{CM}$
necessary to establish an accurate value for the effective static friction $F_R$ will require a fine tuning
of corrugation and applied force. Upon increasing the force however, the instant $F=F_R$ is reached and
the symmetric position $\delta z_{CM}=0$ is attained, the chain will spontaneously jump over the
barrier to the opposite Ising order parameter valley. By e.g., alternating in time rightward and leftward
forces, the effective static friction $F_R$ could become readily detectable
in experiment as the threshold force magnitude for the chain's CM to sway visibly leftward and rightward.
The present static friction study is conducted at zero temperature, ignoring quantum effects, without
external sources of either heating, cooling or damping, and assuming that a sufficiently strong effective
periodic corrugation potential can be laser generated. 
While each of these points requires careful consideration, none, as will be discussed below, is in our view 
fatal to the results presented for static friction.
The proposed dynamic friction experiment would heat up the trapped ion chains, and
cause their destruction once the Joule heating is excessive. We verified however that even at a temperature
of $\sim 0.55$ K, orders of magnitude higher than the effective chain Debye temperature of $\sim 145 \mu$K
(considering Mg$^+$ ions separated by $a_0=5 \mu$m), the chain is still stable in the strong trapping potential.

In conclusion, the field of cold atoms has had a tremendous development, recently covering important interdisciplinary areas
such as Anderson localization~\cite{inguscio}. Now the potential tribological impact of cold ion results could
also be significant. The onset of static friction is a rich problem of current interest~\cite{frenken,hirano2},
and so is the dynamic friction growth with corrugation\cite{coffey}. Most importantly, ion chains represent
a new clean design, free of most complications of real crystal lattices and with many more controllable parameters
for a study of basic nanofrictional properties.

\section{Methods}

{\bf Simulation Details}. The equations of motion derived from (\ref{hamiltonian}) turn dimensionless by defining the
quantity $d^3=e^2/m \omega_{\perp}^2$, and expressing distance in units of $d$, time in units of
$1/\omega_{\perp}$ and the mass in units of $m$. With this definition, energy and force are
also dimensionless in units of $e^2/d$ and $e^2/d^2$ respectively.
Simulations are carried out using classical molecular dynamics, when necessary with a weak frictional damping,
and mostly at zero or low temperature. The equations of motion have been integrated using a velocity-Verlet
algorithm with a time step $\Delta t=5\cdot 10^{-3}$ time units. The threshold trap potential asymmetry
$R_0$ for linear/zig-zag structure transition is determined with adiabatic increments $\Delta R=10^{-4}$
starting from $R=0.5\cdot 10^{-3}$. The critical periodic corrugation potential amplitude $U_c$ for
the symmetric-asymmetric transition of linear and zig-zag chains is obtained by adiabatic increments
$\Delta U_0=10^{-5}$ starting from an initial value $U_0=0.012$ for the zig-zag configuration and $U_0=0.015$
for the linear chain. In all the adiabatic increment procedures a small viscous damping $-\gamma {\bf v}_i$,
with $\gamma=0.01$, is applied to all ions in order to eliminate the excess energy. With our choice
of $\gamma$ a total time of $2500$ between successive increments is enough to damp all the excess energy away.

{\bf Physical parameters and orders of magnitude}

a)Temperature and quantum effects. The present study is conducted conventionally at zero temperature, ignoring quantum effects, without
external sources of heating, cooling or damping, and assuming that a sufficiently strong periodic potential
can be laser generated for the transition to take place at the given incommensurability provided by the laser field.
Even though temperature is in reality finite (typically of order $100$ nK), the previously reported comparison 
between left and right chain distortion energies with $k_BT$ implies a symmetry-broken state lifetime much longer than any 
measurement time, justifiying the zero temperature treatment, where data and their 
significance are clearer. In this regime 
moreover,
quantum zero-point effects could in principle alter some of the details close to the depinning
threshold, but are otherwise not expected to affect the overall friction picture, both static and dynamic.

b)  Corrugation potential magnitude. To produce the effective 
periodic corrugation 
potential, the wavelength $\xi$ of a laser light must fit an electronic excitation in the spectrum of
the chosen ion species (e.g. $280$ nm for Mg$^+$, $729$ nm for Ca$^+$ or $399$ nm for Yb$^+$),
with a typical $a_0=5\mu$m the golden mean incommensurate periodic potential wavelength is $\lambda= 3000 nm \gg \xi$,
nevertheless the desired $\lambda$ value for the effective periodic potential can be obtained
by laser beam crossing at a 
chosen
angle [24].
Experimentally, it is mandatory  to realize a sufficient corrugation 
amplitude $U=U_0$ for the 
Aubry
transition to occur. Assuming for example Mg$^+$ ions
%
and golden ratio incommensurability, 
the critical amplitude of the effective periodic potential $U_c$ requires a laser intensity of some $KW/m^2$, 
which seems entirely within reach.
On the other hand, the Aubry pinning transition becomes, as is well known, easier with 
any chain-corrugation incommensurability different from the golden ratio  [25],
whereby in general the transition would occur with even with weaker corrugation amplitudes than $ U_0$.

c)  Phase misalignment effects. We assumed so far the possibility to position the corrugation potential maximum
precisely at the minimum of the harmonic trap, a feat
which may be experimentally difficult. 
Slight shifts of the corrugation potential phase away from that point would however
simply smear the Aubry pinning onset in the same way a weak symmetry breaking field transition would affect a phase
transition. 

d) Chain heating during ringdown. In the oscillation ringdown simulations, the initial potential energy difference $8.9$, due to the
chain displacement of $10$ $a_0$, is gradually converted to heating of the chain. With the mass of Mg$^+$, the
final temperature reached is approximately $0.55$ K, orders of magnitude higher than the effective chain Debye
temperature of $\sim 145 \mu$K. Despite that, we verified by direct simulation that at this temperature the
chain is still stable in the strong trapping potential.

{\bf Acknowledgements}\\
This work is part of Eurocores Projects FANAS/AFRI sponsored by the Italian Research Council (CNR), 
and of FANAS/ACOF. It is also sponsored in part by the Italian Ministry of University and Research, 
through PRIN/COFIN contracts 20087NX9Y7 and 2008y2p573. ET thanks M. Inguscio for telling him about cold ion traps.

{\bf Author contributions}\\
A.B. and A.V. carried out the MD simulations. E.T. coordinated the work. 
All authors contributed to the numerical data analysis, paper writing and revising.

{\bf Additional information}\\
Supplementary information accompanies this paper on www.nature.com/naturecommunications. 

Correspondence and requests for materials should be addressed to E.T., tosatti@sissa.it.

Competing financial interests: The authors declare no competing financial interests.

Reprints and permissions information is available on-line at
http://npg.nature.com/reprintsandpermissions.


\begin{thebibliography}{10}
\expandafter\ifx\csname url\endcsname\relax
  \def\url#1{\texttt{#1}}\fi
\expandafter\ifx\csname urlprefix\endcsname\relax\def\urlprefix{URL }\fi
\providecommand{\bibinfo}[2]{#2}
\providecommand{\eprint}[2][]{\url{#2}}

\bibitem{hirano}
\bibinfo{author}{Shinjo, K.} \& \bibinfo{author}{Hirano, M.}
\newblock \bibinfo{title}{Dynamics of friction- superlubric state}.
\newblock \emph{\bibinfo{journal}{Surface Science}}
  \textbf{\bibinfo{volume}{283}}, \bibinfo{pages}{473--478}
  (\bibinfo{year}{1993}).

\bibitem{dienwiebel}
\bibinfo{author}{Dienwiebel, M.} \emph{et~al.}
\newblock \bibinfo{title}{Superlubricity of graphite}.
\newblock \emph{\bibinfo{journal}{Phys. Rev. Lett.}}
  \textbf{\bibinfo{volume}{92}}, \bibinfo{pages}{126101}
  (\bibinfo{year}{2004}).

\bibitem{aubry}
\bibinfo{author}{Floria, L.} \& \bibinfo{author}{Mazo, J.}
\newblock \bibinfo{title}{Dissipative dynamics of the frenkel-kontorova model}.
\newblock \emph{\bibinfo{journal}{Advances in Physics}}
  \textbf{\bibinfo{volume}{45}}, \bibinfo{pages}{505} (\bibinfo{year}{1996}).

\bibitem{braun}
\bibinfo{author}{Braun, O.} \& \bibinfo{author}{Kivshar, Y.}
\newblock \emph{\bibinfo{title}{The Frenkel-Kontorova Model: Concepts, Methods,
  and Applications}} (\bibinfo{publisher}{Springer-Verlag},
  \bibinfo{address}{Berlin, Germany}, \bibinfo{year}{2004}).

\bibitem{persson}
\bibinfo{author}{Persson, B.~N.}
\newblock \emph{\bibinfo{title}{Sliding Friction}}
  (\bibinfo{publisher}{Springer-Verlag}, \bibinfo{address}{Berlin, Germany},
  \bibinfo{year}{1998}).

\bibitem{cieplak}
\bibinfo{author}{Cieplak, M.}, \bibinfo{author}{Smith, E.} \&
  \bibinfo{author}{Robbins, M.}
\newblock \bibinfo{title}{Molecular origins of friction - the force on
  adsorbates}.
\newblock \emph{\bibinfo{journal}{Science}} \textbf{\bibinfo{volume}{265}},
  \bibinfo{pages}{1209} (\bibinfo{year}{1994}).

\bibitem{coffey}
\bibinfo{author}{Coffey, T.} \& \bibinfo{author}{Krim, J.}
\newblock \bibinfo{title}{Impact of substrate corrugation on the sliding
  friction levels of adsorbed films}.
\newblock \emph{\bibinfo{journal}{Phys. Rev. Lett.}}
  \textbf{\bibinfo{volume}{95}}, \bibinfo{pages}{076101}
  (\bibinfo{year}{2005}).

\bibitem{ghosh}
\bibinfo{author}{Ghosh, P.}
\newblock \emph{\bibinfo{title}{Ion Traps}} (\bibinfo{publisher}{Claredon
  Press}, \bibinfo{address}{Oxford}, \bibinfo{year}{1995}).

\bibitem{morigi}
\bibinfo{author}{Morigi, G.} \& \bibinfo{author}{Fishman, S.}
\newblock \bibinfo{title}{Eigenmodes and thermodynamics of a coulomb chain in a
  harmonic potential}.
\newblock \emph{\bibinfo{journal}{Phys. Rev. Lett.}}
  \textbf{\bibinfo{volume}{93}}, \bibinfo{pages}{170602}
  (\bibinfo{year}{2004}).

\bibitem{morigi2}
\bibinfo{author}{Morigi, G.} \& \bibinfo{author}{Fishman, S.}
\newblock \bibinfo{title}{Dynamics of an ion chain in a harmonic potential}.
\newblock \emph{\bibinfo{journal}{Phys. Rev. E}} \textbf{\bibinfo{volume}{70}},
  \bibinfo{pages}{066141} (\bibinfo{year}{2004}).

\bibitem{raizen}
\bibinfo{author}{Raizen, M.~G.}, \bibinfo{author}{Gilligan, J.~M.},
  \bibinfo{author}{Bergquist, J.~C.}, \bibinfo{author}{Itano, W.~M.} \&
  \bibinfo{author}{Wineland, D.~J.}
\newblock \bibinfo{title}{Ionic crystals in a linear paul trap}.
\newblock \emph{\bibinfo{journal}{Phys. Rev. A}} \textbf{\bibinfo{volume}{45}},
  \bibinfo{pages}{6493--6501} (\bibinfo{year}{1992}).

\bibitem{schiffer}
\bibinfo{author}{Schiffer, J.~P.}
\newblock \bibinfo{title}{Phase transitions in anisotropically confined ionic
  crystals}.
\newblock \emph{\bibinfo{journal}{Phys. Rev. Lett.}}
  \textbf{\bibinfo{volume}{70}}, \bibinfo{pages}{818} (\bibinfo{year}{1993}).

\bibitem{fishman}
\bibinfo{author}{Fishman, S.}, \bibinfo{author}{De~Chiara, G.},
  \bibinfo{author}{Calarco, T.} \& \bibinfo{author}{Morigi, G.}
\newblock \bibinfo{title}{Structural phase transitions in low-dimensional ion
  crystals}.
\newblock \emph{\bibinfo{journal}{Phys. Rev. B}} \textbf{\bibinfo{volume}{77}},
  \bibinfo{pages}{064111} (\bibinfo{year}{2008}).

\bibitem{kienle}
\bibinfo{author}{Kienle, P.}
\newblock \bibinfo{title}{Sunshine by cooling}.
\newblock \emph{\bibinfo{journal}{Naturwissenschaften}}
  \textbf{\bibinfo{volume}{88}}, \bibinfo{pages}{313} (\bibinfo{year}{2001}).

\bibitem{Hermanspahn}
\bibinfo{author}{Hermanspahn, N.} \emph{et~al.}
\newblock \bibinfo{title}{Observation of the continuous stern-gerlach effect on
  an electron bound in an atomic ion}.
\newblock \emph{\bibinfo{journal}{Phys. Rev. Lett.}}
  \textbf{\bibinfo{volume}{84}}, \bibinfo{pages}{427} (\bibinfo{year}{2000}).

\bibitem{Koerber}
\bibinfo{author}{Koerber, T.~W.}, \bibinfo{author}{Schacht, M.~H.},
  \bibinfo{author}{Hendrickson, K. R.~G.}, \bibinfo{author}{Nagourney, W.} \&
  \bibinfo{author}{Fortson, E.~N.}
\newblock \bibinfo{title}{rf spectroscopy with a single ba+ ion}.
\newblock \emph{\bibinfo{journal}{Phys. Rev. Lett.}}
  \textbf{\bibinfo{volume}{88}}, \bibinfo{pages}{143002}
  (\bibinfo{year}{2002}).

\bibitem{Bize}
\bibinfo{author}{Bize, S.} \emph{et~al.}
\newblock \bibinfo{title}{Testing the stability of fundamental constants with
  the hg-199(+) single-ion optical clock}.
\newblock \emph{\bibinfo{journal}{Phys. Rev. Lett.}}
  \textbf{\bibinfo{volume}{90}}, \bibinfo{pages}{150802}
  (\bibinfo{year}{2003}).

\bibitem{Poitzsch}
\bibinfo{author}{Poitzsch, M.~E.}, \bibinfo{author}{Bergquist, J.~C.},
  \bibinfo{author}{Itano, W.~M.} \& \bibinfo{author}{Wineland, D.~J.}
\newblock \bibinfo{title}{Cryogenic linear ion trap for accurate spectroscopy}.
\newblock \emph{\bibinfo{journal}{Rev. Sci. Instrum.}}
  \textbf{\bibinfo{volume}{67}}, \bibinfo{pages}{129} (\bibinfo{year}{1996}).

\bibitem{Molhave}
\bibinfo{author}{Molhave, K.} \& \bibinfo{author}{Drewsen, M.}
\newblock \bibinfo{title}{Formation of translationally cold mgh+ and mgd+
  molecules in an ion trap}.
\newblock \emph{\bibinfo{journal}{Phys. Rev. A}} \textbf{\bibinfo{volume}{62}},
  \bibinfo{pages}{011401(R)} (\bibinfo{year}{2000}).

\bibitem{Cirac}
\bibinfo{author}{Cirac, J.~I.} \& \bibinfo{author}{Zoller, P.}
\newblock \bibinfo{title}{Quantum computation with cold trapped ions}.
\newblock \emph{\bibinfo{journal}{Phys. Rev. Lett.}}
  \textbf{\bibinfo{volume}{74}}, \bibinfo{pages}{4091} (\bibinfo{year}{1995}).

\bibitem{Steane}
\bibinfo{author}{Steane, A.}
\newblock \bibinfo{title}{The ion trap quantum information processor}.
\newblock \emph{\bibinfo{journal}{Appl. Phys. B: Lasers Opt.}}
  \textbf{\bibinfo{volume}{64}}, \bibinfo{pages}{623} (\bibinfo{year}{1997}).

\bibitem{Schmidt}
\bibinfo{author}{Schmidt-Kaler, F.} \emph{et~al.}
\newblock \bibinfo{title}{Realization of the cirac-zoller controlled-not
  quantum gate}.
\newblock \emph{\bibinfo{journal}{Nature}} \textbf{\bibinfo{volume}{422}},
  \bibinfo{pages}{408} (\bibinfo{year}{2003}).

\bibitem{Leibfried}
\bibinfo{author}{Leibfried, D.} \emph{et~al.}
\newblock \bibinfo{title}{Experimental demonstration of a robust, high-fidelity
  geometric two ion-qubit phase gate}.
\newblock \emph{\bibinfo{journal}{Nature}} \textbf{\bibinfo{volume}{422}},
  \bibinfo{pages}{412} (\bibinfo{year}{2003}).

\bibitem{mata}
\bibinfo{author}{Garcia-Mata, I.}, \bibinfo{author}{Zhirov, O.} \&
  \bibinfo{author}{Shepelyansky, D.}
\newblock \bibinfo{title}{Frenkel-kontorova model with cold trapped ions}.
\newblock \emph{\bibinfo{journal}{Eur. Phys. J. D}}
  \textbf{\bibinfo{volume}{41}}, \bibinfo{pages}{325} (\bibinfo{year}{2007}).

\bibitem{mukamel}
\bibinfo{author}{Biham, O.} \& \bibinfo{author}{Mukamel, D.}
\newblock \bibinfo{title}{Global universality in the frenkel-kontorova model}.
\newblock \emph{\bibinfo{journal}{Phys. Rev. A}} \textbf{\bibinfo{volume}{39}},
  \bibinfo{pages}{5326--5335} (\bibinfo{year}{1989}).

\bibitem{sharma}
\bibinfo{author}{Sharma, S.~R.}, \bibinfo{author}{Bergersen, B.} \&
  \bibinfo{author}{Joos, B.}
\newblock \bibinfo{title}{Aubry transition in a finite modulated chain}.
\newblock \emph{\bibinfo{journal}{Phys. Rev. B}} \textbf{\bibinfo{volume}{29}},
  \bibinfo{pages}{6335--6340} (\bibinfo{year}{1984}).

\bibitem{furuya}
\bibinfo{author}{Furuya, K.} \& \bibinfo{author}{de~Almeida, A.}
\newblock \bibinfo{title}{Soliton energies in the standard map beyond the
  chaotic threshold}.
\newblock \emph{\bibinfo{journal}{J. Phys. A}} \textbf{\bibinfo{volume}{20}},
  \bibinfo{pages}{6211} (\bibinfo{year}{1987}).

\bibitem{braiman}
\bibinfo{author}{Braiman, Y.}, \bibinfo{author}{Baumgarten, J.},
  \bibinfo{author}{Jortner, J.} \& \bibinfo{author}{Klafter, J.}
\newblock \bibinfo{title}{Symmetry-breaking transition in finite
  frenkel-kontorova chains}.
\newblock \emph{\bibinfo{journal}{Phys. Rev. Lett.}}
  \textbf{\bibinfo{volume}{65}}, \bibinfo{pages}{2398--2401}
  (\bibinfo{year}{1990}).

\bibitem{coppersmith}
\bibinfo{author}{Coppersmith, S.~N.} \& \bibinfo{author}{Fisher, D.~S.}
\newblock \bibinfo{title}{Threshold behavior of a driven incommensurate
  harmonic chain}.
\newblock \emph{\bibinfo{journal}{Phys. Rev. A}} \textbf{\bibinfo{volume}{38}},
  \bibinfo{pages}{6338--6350} (\bibinfo{year}{1988}).

\bibitem{peyrard}
\bibinfo{author}{Peyrard, M.} \& \bibinfo{author}{Aubry, S.}
\newblock \bibinfo{title}{Critical-behavior at the transition by breaking of
  analyticity in the discrete frenkel-kontorova model}.
\newblock \emph{\bibinfo{journal}{J. Phys. C: Solid State Phys.}}
  \textbf{\bibinfo{volume}{16}}, \bibinfo{pages}{1593} (\bibinfo{year}{1983}).

\bibitem{braun_paper}
\bibinfo{author}{Braun, O.~M.}, \bibinfo{author}{Bishop, A.~R.} \&
  \bibinfo{author}{R\"oder, J.}
\newblock \bibinfo{title}{Hysteresis in the underdamped driven
  frenkel-kontorova model}.
\newblock \emph{\bibinfo{journal}{Phys. Rev. Lett.}}
  \textbf{\bibinfo{volume}{79}}, \bibinfo{pages}{3692--3695}
  (\bibinfo{year}{1997}).

\bibitem{consoli}
\bibinfo{author}{Consoli, L.}, \bibinfo{author}{Knops, H.} \&
  \bibinfo{author}{Fasolino, A.}
\newblock \bibinfo{title}{Onset of sliding friction in incommensurate systems}.
\newblock \emph{\bibinfo{journal}{Phys. Rev. Lett.}}
  \textbf{\bibinfo{volume}{85}}, \bibinfo{pages}{302} (\bibinfo{year}{2000}).

\bibitem{liebsch}
\bibinfo{author}{Liebsch, A.}, \bibinfo{author}{Goncalves, S.} \&
  \bibinfo{author}{Kiwi, M.}
\newblock \bibinfo{title}{Electronic versus phononic friction of xenon on
  silver}.
\newblock \emph{\bibinfo{journal}{Phys. Rev. B}} \textbf{\bibinfo{volume}{60}},
  \bibinfo{pages}{5034} (\bibinfo{year}{1999}).

\bibitem{inguscio}
\bibinfo{author}{Aspect, A.} \& \bibinfo{author}{Inguscio, M.}
\newblock \bibinfo{title}{Anderson localization of ultracold atoms}.
\newblock \emph{\bibinfo{journal}{Physics Today}}
  \textbf{\bibinfo{volume}{62}}, \bibinfo{pages}{30} (\bibinfo{year}{2009}).

\bibitem{frenken}
\bibinfo{author}{Dienwiebel, M.} \emph{et~al.}
\newblock \bibinfo{title}{Superlubricity of graphite}.
\newblock \emph{\bibinfo{journal}{Phys. Rev. Lett.}}
  \textbf{\bibinfo{volume}{92}}, \bibinfo{pages}{126101}
  (\bibinfo{year}{2004}).

\bibitem{hirano2}
\bibinfo{author}{Hirano, M.}
\newblock \bibinfo{title}{Atomistics of friction}.
\newblock \emph{\bibinfo{journal}{Surface Science Reports}}
  \textbf{\bibinfo{volume}{60}}, \bibinfo{pages}{159} (\bibinfo{year}{2006}).

\end{thebibliography}
\end{document}